\documentclass[a4paper,reqno,11pt]{amsart}

\usepackage{amsmath,amsthm,amssymb,color,graphicx}
\usepackage{amscd,verbatim}
\usepackage[all]{xy}

\theoremstyle{plain}
\newtheorem{theorem}{Theorem}

\theoremstyle{definition}
\newtheorem{definition}[theorem]{Definition}

\theoremstyle{remark}
\newtheorem*{remark}{Remark}

\numberwithin{equation}{section}
\numberwithin{theorem}{section}
\numberwithin{figure}{section}
\numberwithin{table}{section}

\newcommand{\cL}{{\mathcal L}}

\newcommand{\RR}{{\mathbb R}}
\newcommand{\TT}{{\mathbb T}}
\newcommand{\ZZ}{{\mathbb Z}}

\newcommand{\dwedge}{{\sqcap}}

\newcommand{\fg}{{\mathfrak g}}
\newcommand{\ft}{{\mathfrak t}}

\newcommand{\sfG}{{\mathsf G}}
\newcommand{\sfT}{{\mathsf T}}

\begin{document}

\title{T-duality for principal torus bundles}

\author[P Bouwknegt]{Peter Bouwknegt}

\address[Peter Bouwknegt]{
Department of Physics and Mathematical Physics, 
and Department of Pure Mathematics \\
University of Adelaide \\
Adelaide, SA 5005 \\
Australia}
\email{pbouwkne@physics.adelaide.edu.au, 
       pbouwkne@maths.adelaide.edu.au}

\author[K Hannabuss]{Keith Hannabuss}

\address[Keith Hannabuss]{
Department of Mathematics\\
University of Oxford\\
England}
\email{kch@balliol.oxford.ac.uk}

\author[V Mathai]{Varghese Mathai}

\address[Varghese Mathai]{
Department of Pure Mathematics \\
University of Adelaide \\
Adelaide, SA 5005 \\
Australia}
\email{vmathai@maths.adelaide.edu.au}

\thanks{PB and VM are financially supported by the Australian Research Council, and
KH would like to thank the Department of Pure Mathematics at the University of Adelaide
for hospitality during the course of this project.}

\begin{abstract}
In this paper we study T-duality for principal torus bundles with H-flux.  We identify 
a subset of fluxes which are T-dualizable, and compute both the dual
torus bundle as well as the dual H-flux.  We briefly discuss the generalized Gysin
sequence behind this construction and provide examples both of non T-dualizable
and of T-dualizable H-fluxes.
\end{abstract}
\maketitle

\section{Introduction}

T-duality is one of the most powerful tools in (super)string theory.  It provides
an equivalence between string theories which, in their low-energy 
field theory limit might superficially look very different, but are in fact the same
in the sense that there exists a 1--1 correspondence between fields, states, etc.
In particular, T-duality should relate the various D-branes in the theory, which 
is mathematically expressed as the fact that there should be an isomorphism 
between the relevant (twisted) K-theories (and their close cousin, (twisted)
cohomology).

Locally, the T-duality transformation rules on the massless fields in string theory,
known as the Buscher  rules, have been known for quite some time
\cite{Bus}.  Global issues, though, in particular in the background of NS H-flux ,
have remained obscure (see \cite{AABL} for some early investigations).

In a recent paper, T-duality for principal circle bundles $\pi:E\to M$ (i.e.\ circle 
bundles with a free circle action), in the background
of H-flux $[H]\in H^3(E,\ZZ)$ was examined \cite{BEMa,BEMb}.  Such bundles are
classified by their first Chern class $c_1(E)\in H^2(M,\ZZ)$, and 
it was shown that T-duality interchanges the fibrewise integral of the H-flux
with the first Chern class.  I.e.\ $(E,H)$ and its T-dual $(\widehat E,\widehat H)$
are related by 
\begin{equation}
  c_1(E) = \int_{\widehat \TT} \ \widehat H \,,\qquad
  c_1(\widehat E) = \int_\TT \ H \,,
\end{equation}
as can easily be argued from the Gysin sequences of the bundles $E$
and $\widehat E$.
In addition, the isomorphisms between the twisted cohomologies 
and twisted K-theories
of $(E,H)$ and its T-dual $(\widehat E,\widehat H)$ were explicitly
constructed.

In the present paper we will generalize these results to principal torus bundles,
and identify a subset of {\it T-dualizable} H-fluxes which admit T-duals in a 
completely analogous way to the circle bundle case.  
We certainly do not want to claim that T-duals do not exist if one
goes outside of this restrictive class, in fact it is well-known that the torus action
need not be free, i.e. the torus action may have fixed points.  Many examples
of T-duals in this more general set-up, in particular in the context of mirror symmetry 
(see, e.g., \cite{SYZ} for the original idea, and \cite{DP} for most recent developments), 
have been constructed, but as far as we know no complete picture is known in 
the presence of H-flux.  The purpose of this paper was to identify a class of 
torus bundles and H-fluxes which admit T-duals in the same class, and lead to 
isomorphisms in twisted
cohomology and twisted K-theory, in a manner which generalizes the case of circle 
bundles.  In particular we were aiming for a (generalized) Gysin sequence
which relates the cohomologies of the torus-bundle $(E,H)$ and its T-dual 
$(\widehat E,\widehat H)$.

The restriction to principal torus bundles is a natural one, physically it 
corresponds to the situation where momentum along the torus directions
is conserved.  In the case of circles bundles, all orientable circle bundles are
in fact principal circle bundles.  An example of a non-principal (and therefore
non-orientable)  circle bundle 
over the circle is the Klein bottle for which the analysis in \cite{BEMa} does not apply.
For torus bundles, though, the situation is completely different.  There are
many more torus bundles than principal torus bundles (principal torus
bundles over the torus have been classified in \cite{PS}).  Moreover, even in the
case of principal torus bundles not all H-fluxes admit a straightforward T-dual.
There is a subset of H-fluxes, essentially those closed 3-forms which only have
one `leg' in the torus direction, which do however admit a T-dual, which is again
a principal torus bundle with an H-flux in the same `T-dualizable' subset.

The paper is organized as follows.  In Section \ref{secB} we discuss the subset
of T-dualizable H-fluxes on principal torus bundles, and give an explicit characterization
of both the dual torus bundle as well as the dual H-flux.  We also show how this would fit
into a generalized Gysin sequence.  In Section \ref{secC} we discuss some examples
of both T-dualizable torus bundles (such as the group manifold) and non T-dualizable
torus bundles, and discuss the complications which arise for non principal torus 
bundles.

For reasons of clarity, we restrict the discussion in this paper to T-duality 
aspects pertaining to the image of integral cohomology classes in
de-Rham cohomology, i.e.\ the cohomology of differential 
forms with integral periods.  
The full result, as well as further details on the results in this paper,
will be dealt with in a companion paper \cite{BHMb}.

\section{T-duality for principal torus bundles}  \label{secB}

\subsection{$T$-dualizable H-fluxes} \label{secBA}

Let us denote the circle by $\TT$, the $n$-torus by $\TT^n$,
or $\sfT$ for short.  $\sfT$ can be 
considered as an (abelian) Lie group, and we let
$\ft$ denote the Lie algebra of $\sfT$, and $\hat\ft$ the
dual Lie algebra.

Now let $\pi: E\to M$ be a principal $\sfT$-bundle.  
The action of $\sfT$ on $E$ associates to each element $X\in\ft$
a vector field on $E$ which, by abuse of notation, we will also
denote as $X$.  We will denote the Lie derivative and contraction 
with respect to the vector field $X$ as $\cL_X$ and $\imath_X$, respectively.

Let $\Omega^k(N)$ and $\Omega^k(N,\ft)$ denote the set of $k$-forms,
and $\ft$-valued $k$-forms, on
$N$, respectively, and let $H^k(N)$ and $H^k(N,\ft)$ be the associated
de-Rham cohomology groups of differential forms with integral periods.
[In the rest of the paper the integrality conditions on closed forms will not be explicitly 
stated.]
A form $\omega \in \Omega^k(E)$ is called {\it basic} if $\omega$ is the pull-back
of a form on the base manifold $M$.  This is equivalent to the requirement
that $\cL_X \omega = \imath_X\omega = 0$ for all $X\in\ft$.
An H-flux on $E$ is, by definition, a closed, integral, 3-form $H\in\Omega^3(E)$, i.e.
it determines a class $[H]\in H^3(E)$.

\label{defB}
\begin{definition} An H-flux $H$ is called {\it T-dualizable} when there exists
a closed, $\hat\ft$-valued, 2-form $\widehat F$ on $M$ such that
the pair $(H,\widehat F)$ satisfies
\begin{equation} \label{eqBa}
dH = 0 \,,\qquad \imath_X H = \pi^*\widehat F(X)\,,
\end{equation}
for all $X\in\ft$, where $\widehat F(X)\in \Omega^2(M)$ 
denotes the dual pairing of
$\widehat F\in \Omega^2(M,\hat \ft)$ with $X\in \ft$.
\end{definition}

\begin{remark}
More generally, one may define a T-dualizable flux as a pair $(H,\widehat F)$
such that the relations \eqref{eqBa} hold at the level of cohomology only.
As this would unnecessarily complicate
the discussion below, we will simply assume that representatives $(H,\widehat F)$
have been chosen such that \eqref{eqBa} holds at the level of forms.
\end{remark}

Note, that the closed 2-form $\widehat F$ on $M$, i.e. 
$[\widehat F] \in H^2(M,\hat \ft)$, determines a 
principal $\widehat \sfT$-bundle $\hat\pi: \widehat E \to M$, which we will refer
to as the T-dual torus bundle.  In fact, to be precise, $[\widehat F]$ only determines
$\widehat E$ up to torsion.  To determine the torsion part of $\widehat E$ we need 
to work with integer cohomology classes.  As the purpose of this paper is to explain the 
main ideas behind the construction,  we will simply accept the fact that if we were to work
with the `appropriate' cohomology theory the T-dual bundle $\widehat E$ would be
uniquely determined (up to isomorphism).  As far as providing the isomorphism of 
the twisted cohomology of $E$ with that of $\widehat E$ the torsion part of $\widehat E$ is
irrelevant.

{}From Definition \ref{defB} it follows that all T-dualizable H-fluxes
$(H,\widehat F)$ necessarily satisfy
$\cL_X H = 0 = \cL_X \widehat F$ for all $X\in\ft$.
In fact, it is well-known that every closed form on $E$ is cohomologous to a 
closed form $\omega$
on $E$ that satisfies $\cL_X \omega=0$. So, without loss of generality, we may assume
that all forms in question are invariant.
On this subspace the de-Rham differential $d$ anti-commutes with the contraction $\imath_X$,
since $\cL_X = \{ d, \imath_X\}$, so, after defining a locally defined 
1-form $\widehat A\in\Omega_{\text{loc}}^1(E,\ft)$ such that 
$\pi^*\widehat F = d\widehat A$, we may 
interpret the conditions \eqref{eqBa}
on the pair $(H,\widehat A)$ as defining some sort of Deligne cohomology 
class in a double complex, except
for the fact that $\widehat A$ is only locally defined on $E$. 
One might be tempted to think that the pull-back $\pi^*\widehat F$ is exact on $E$,
and that therefore $\widehat A$ is  
globally defined on $E$, but this is incorrect.  In fact, the set-up is precisely
such that $\hat\pi^* \widehat F$ is exact  on the T-dual bundle 
$\widehat E$, and that the various locally defined $\widehat A$ patch together to
form a globally defined connection 1-form, with values in $\hat\ft$, on $\widehat E$,
such that $\hat\pi^* \widehat F = d\widehat A$.  Without loss of generality we assume
that $\widehat A\in\Omega^1(\widehat E,\hat\ft)$ is normalized such that 
\begin{equation} \label{eqBb}
\imath_X \widehat A = X \,,
\end{equation}
for all $X\in\hat\ft$.

\begin{remark}
Finally, we remark that the conditions \eqref{eqBa}, written in terms of $(H,\widehat A)$,
are remarkably similar to the conditions defining an equivariant cohomology class,
i.e., a class in $H^3_\sfT(E)$.  Equivariant cohomology works with differential forms
which are also polynomials on the Lie algebra $\ft$ of the group $\sfT$, where the total
degree of a homogeneous form $\alpha$ is defined as the sum of the form degree of
$\alpha$ plus twice the polynomial degree.  The equivariant differential $d_\sfT$ is 
defined as
\begin{equation}
d_\sfT \alpha(X) = d\alpha(X) - \imath_X \alpha(X) \,.
\end{equation}
Written in terms of $\alpha=H + \widehat A$, the condition that $\alpha$ is closed under
$d_\sfT$ gives, after collecting the terms of the same form degree,
\begin{equation}
dH=0 \,,\qquad \imath_X H = d\widehat A(X)\,,\qquad
\imath_X \widehat A(X) = 0 \,.
\end{equation}
The first two equations correspond to \eqref{eqBa}, while the latter is a normalization
condition on $\widehat A$.
The difference between Eqn.~\eqref{eqBa} and equivariant cohomology is, though,
that $\widehat A$ is only defined locally.  In fact, since $\sfT$ acts freely on $E$,
one would have $H_\sfT(E) \cong H(M)$, which makes equivariant cohomology 
not particularly useful in this case.
\end{remark}

\subsection{The T-dual H-flux} \label{secBB}

In Sect.~\ref{secBA} we have seen how,
given a $\sfT$-bundle $\pi:E\to M$, and a T-dualizable H-flux $(H,\widehat F)$,
we have defined a T-dual $\widehat \sfT$-bundle $\hat\pi : \widehat E \to M$, 
with connection $\widehat A \in \Omega^1(\widehat E,\hat\ft)$, such that
$\hat \pi^* \widehat F = d\widehat A$, and normalized according to Eqn.~\eqref{eqBb}.
The goal of this section is to define the T-dual H-flux.  It will turn out
that the T-dual $\widehat H$ is a T-dualizable H-flux on $\widehat E$,
as one would hope.

As in \cite{BEMb} we could proceed to find $\widehat H$, up to a basic form,
from a generalized Gysin sequence (see Sect.~\ref{secBC}).  Here we proceed
by simply defining $\widehat H$ and show it has the required properties.

Let $A$ be a connection 1-form on $E$, that is $A\in \Omega^2(E,\ft)$
such that its curvature $F = dA$ is (the pull-back of)
a closed 2-form on $M$.  We normalize $A$ such that
$\imath_X A = X$.  These together imply that $\cL_X A=0$.
Since the T-dual connection and H-flux live on a different space as the original 
connection and H-flux, in order to compare them we need to pull all of these forms
back to a common space,
known as a {\it correspondence space}.  The correspondence space, in this case, is
the fibred product of $E$ and $\widehat E$, i.e.
\begin{equation}
E \times_M \widehat E = \{ (x,\hat x) \in E\times\widehat E\ | \ \pi(x) = \hat\pi(\hat x)\} \,.
\end{equation}
The projection  $p: E \times_M \widehat E\to E$ is given as the composition 
\begin{equation} \begin{CD}
E\times_M \widehat E @>1\otimes \hat \pi>> E\times_M M @>\cong>> E
\end{CD}
\end{equation}
and defines $E\times_M\widehat E$ as a principal $\widehat\sfT$-bundle
over $E$.  
Similarly for $\hat p$.  This shows we have a commutative diagram of
torus bundles
\begin{equation*} 
\xymatrix @=8pc @ur { E \ar[d]_{\pi} &
E\times_M  \widehat E \ar[d]_{\hat p} \ar[l]^{p} \\ M & \widehat  
E\ar[l]^{\hat \pi}}
\end{equation*}
In the remainder of this section all forms are pulled back to the space on which the
equation makes sense,
but for notational simplicity we will omit the pull-backs from the equations.

First of all, consider the difference 
$\Omega = A \dwedge d\widehat A - H$,
where we define the $\dwedge$ as the wedge between
forms followed by the canonical pairing between $\hat \ft$ and $\ft$.
A priori, $\Omega$ is a form on $E$, but we will show that $\Omega$ is 
actually a basic 3-form.
Obviously we have $\cL_X \Omega =0$, and a little calculation gives
\begin{equation*}
\imath_X \Omega = d\widehat A(X) - \imath_X H = 0 \,,
\end{equation*}
which together imply that $\Omega$ is basic.  In other words, this implies
that our T-dualizable H-flux $H$ can be written as 
\begin{equation}
H  = A \dwedge d\widehat A - \Omega \,,
\end{equation}
for some $\Omega\in\Omega^3(M)$.
Note that, in particular, 
\begin{equation}
dA \dwedge d\widehat A = d\Omega\,,
\end{equation}
hence $[F\dwedge\widehat F] =0$ in $H^4(M)$.

We now define the T-dual H-flux on $\widehat E$ to be
\begin{equation}
{\widehat H} = dA \dwedge {\widehat A} - \Omega\,.
\end{equation}
Then, a simple calculation shows that  ${\widehat H}$ is a closed,
$\widehat \sfT$-invariant, 3-form on $\widehat E$.  Moreover it is T-dualizable, since
\begin{equation} 
\imath_{\hat X} \widehat H = dA(\widehat X) = F(\widehat X) \,.
\end{equation}

Finally, observe that on the correspondence space $E \times_M \widehat  E$, 
the difference $H-\widehat H$ is exact\footnote{A similar conclusion was reached in
\cite{Alv}, Sect.~4.}, since 
\begin{equation}
H - \widehat H = A \dwedge d{\widehat A} - dA \dwedge {\widehat A} =  -  
d(A \dwedge {\widehat A}) \,.
\end{equation}
This, in particular implies that $A\dwedge\widehat A$ intertwines between the
twisted cohomologies (and twisted K-theories) of $(E,H)$ and $(\widehat E,\widehat H)$.
The explicit isomorphism between the $\ZZ_2$-graded twisted cohomologies
of $E$ and $\widehat E$, i.e. $H^{\overline i}(E,[H]) \cong H^{\overline{i+n}}(\widehat E,
[\widehat H])$ where $n$ is the rank of the torus and $\overline i = i\mod 2$,
is a straightforward generalization of the result in \cite{BEMa}, namely that the Buscher rules on the
RR fields can be encoded in the formula,
\begin{equation}
G \mapsto \widehat G = \int_{\sfT} \ e^{A\dwedge \widehat A}\ G \,.
\end{equation}
This reduces to the familiar formula for the Buscher rules in \cite{Bus, Hor}, both
locally, and globally when the flux $H=dB$ is exact.
A similar generalization exist for the isomorphism between the respective
twisted K-theories, $K^{\overline i}(E,[H]) \cong K^{\overline{i+n}}(\widehat E,
[\widehat H])$, and the two isomorphisms are compatible in the 
sense that the Riemann-Roch theorem holds in this case as well
(see \cite{BHMb} for more details).

\begin{remark}
It is well-known that an $n$-torus, for $n>1$, admits a group of T-dualities, 
namely $SO(n,n;\ZZ)$.  
Additional T-dualities can be recovered in the formalism above by 
taking non-canonical pairings between $\ft\cong \RR^n$ and $\hat\ft\cong \RR^n$.
\end{remark}

\subsection{The generalized Gysin sequence} \label{secBC}

In this section we will briefly indicate how the above construction would nicely 
fit in the framework of a generalized Gysin sequence.  In fact, the 
existence of this Gysin sequence motivated our restriction to principal
torus bundles and our definition of T-dualizable H-fluxes.
The Gysin sequence we are about to discuss is the de-Rham 
analogue of the Gysin sequence in \cite{PRW}.

In general, a torus bundle, or any fibre bundle, gives
rise to a spectral sequence (the so-called
Leray spectral sequence) computing the cohomology of the bundle space
from the cohomology of the base space and the fibre.  In the case
of sphere bundles this spectral sequence collapses into a long exact
sequence, the Gysin sequence, in cohomology (see, e.g., \cite{BT}),
but this is not the case for torus bundles.  But even if it did, with the
application to T-duality in mind we are interested in relations between 
cohomology groups involving the cohomology group that classifies 
(a subclass of) torus bundles.  Principal $n$-torus bundles over a base 
space $M$ are classified 
by the sheaf cohomology group $H^1(M, \underline\sfT) \cong 
H^2(M,\ZZ^n)$, whose image in de-Rham cohomology can be identified
with $H^2(M,\ft)$.  Thus, for a given torus bundle $\pi:E\to M$ we are
looking for a long exact sequence relating the cohomologies 
$H^k(E)$ and $H^k(M,\ft)$.  The missing ingredient is a third cohomology
group $H_{\text RW}^k(E,\ft)$.  It was introduced, in the sheaf
language, by Raeburn and Williams 
in \cite{RWc}, where it was refered to as $\ft$-equivariant cohomology.  However,
while it is closely related to the conventional equivariant cohomology 
corresponding to the $\sfT$-action on $E$, it is in general not the same
(cf.\ the discussion in Sect.~\ref{secBB}).  To avoid
confusion we will refer to it as the RW-cohomology, and use the notation
$H_{\text RW}^k(E,\ft)$.  Its definition is generalizing Definition \ref{defB}.

An element in $H_{\text RW}^k(E,\ft)$ is a pair $(H,\widehat F)$, with $H\in \Omega^k(E)$
and $\widehat F\in \Omega^{k-1}(M,\hat\ft)$, such that the following 
conditions are satisfied for all $X\in\ft$
\begin{equation}
dH=0\,,\qquad
\imath_X H = \pi^* \widehat F(X)\,,\qquad
d\widehat F=0 \,.
\end{equation}
The RW-cohomology is the set of such pairs modulo pairs of the 
form $(H,\widehat F) = (dB, dC)$, where
$B\in\Omega^2(E)$ and $C\in\Omega^1(M,\hat\ft)$, such that $\imath_X B=\pi^* C(X)$,
for all $X\in\ft$.

As an aside, let us remark that the \v Cech analogue of $H^k_{\text RW}(E,\ft)$ is the sheaf 
cohomology group $H^{k-1}_{\text RW}(E,\ft,\underline\TT)$, and that 
$H^1_{\text RW}(E,\ft,\underline\TT)$ is in 1--1 correspondence with the isomorphism
classes of $\sfT$-equivariant line bundles over $E$ that are locally trivial over
$M=E/\sfT$ \cite{PRW}, while $H^2_{\text RW}(E,\ft,\underline\TT)$ is in 1--1 correspondence 
with the stable isomorphism
classes of $\sfT$-equivariant bundle gerbes  $L \to E^{[2]}$ that are locally trivial over
$E^{[2]}/\sfT$ \cite{BHMb}.

Now, given a principal $\sfT$-bundle $\pi : E\to M$ characterized by
a curvature $F$, i.e.\ $[F]\in H^2(M,\ft)$, we expect to derive a long exact 
sequence in cohomology (for $k\geq1$), the so-called generalized Gysin
sequence, given as in \cite{PRW,BHMb} by,
\begin{equation*}
\begin{CD}
\ldots @>>> H^k(M) @>p^*>> H^k_{\text RW}(E,\ft) @>b>> 
H^{k-1}(M,\hat\ft) @>\dwedge F>> H^{k+1}(M) @>>> \ldots
\end{CD}
\end{equation*}
where the various maps are explicitly given by\footnote{We write the maps 
as given on representatives of the various cohomology groups for notational
simplicity.}
\begin{align}
p^* & : H^k(M) \to H^k_{\text RW}(E,\ft) \,,\qquad   & H  \mapsto (\pi^* H,0) \,,\nonumber \\
b & : H^k_{\text RW}(E,\ft) \to H^{k-1}(M,\hat\ft) \,,\qquad & (H,\widehat F)  \mapsto \widehat F\,,
\nonumber \\
\dwedge F & : H^{k-1}(M,\hat\ft) \to H^{k+1}(M) \,, \qquad & \widehat F  \mapsto
\widehat F\dwedge F \,.
\end{align}
In the last definition, we remind the reader that 
$\widehat F\dwedge F$ stands for taking both the wedge product of the 
$\hat\ft$-valued $(k-1)$-form
$\widehat F$ with the $\ft$-valued 2-form $F$, as well as the canonical
pairing between $\hat\ft$ and
$\ft$, to produce an $\RR$-valued  $(k+1)$-form.

The results and computations in Sects.~\ref{secBA} and \ref{secBB} can easily 
be interpreted as `diagram chasing' in this generalized Gysin sequence (the 
$k=3$ segment, in particular) in an analogous manner to the discussion
in \cite{BEMa}.
Finally, we remark that the circle bundle case is obtained from the 
more general case above by observing that $H^k_{\text RW}(E,\ft) \cong
H^k(E)$, for $k\geq1$, if $E$ is a principal circle bundle (cf.\ \cite{PRW}).

\section{Examples} \label{secC}

\subsection{The trivial $\TT^2$-bundle over $\TT$}  \label{secCA}

There are no nontrivial principal $\TT^2$-bundles over $\TT$ as
$H^1(\TT,\underline{\TT}^2) \cong H^2(\TT,\ZZ^2) =0$.  In particular,
the nilmanifold (`twisted 3-torus') that enters in many physically interesting
examples, is a nontrivial $\TT^2$-bundle over $\TT$, but it is {\it not} 
principal.  

A simple example of a principal torus bundle with non T-dualizable H-flux
is provided by $\TT^3$, considered as the trivial $\TT^2$-bundle over $\TT$, 
with $H$ given by $k$ times the volume form on $\TT^3$.  In this case
$\imath_X H$, with $X\in \ft$, is not a basic 2-form for obvious reasons,
hence $H$ is non T-dualizable in the sense of Definition \ref{defB}.

It is illuminating to work out explicitly what happens in this case, as naively 
one might T-dualize one circle at a time by applying the Buscher  rules (see also
the discussion in \cite{BEMa}).
Explicitly, in terms of Cartesian coordinates $(x,y,z)\sim (x+1,y,z)\sim (x,y+1,z)\sim
(x,y,z+1)$ we have a metric $g$ and H-flux $H$ given by
\begin{equation}
  g = dx^2 + dy^2 + dz^2 \,,\qquad
  H= k\, dx\wedge dy\wedge dz\,.
 \end{equation}
 After choosing a local gauge $H=dB$, with $B=k x\, dy\wedge dz$, application
 of the Buscher  rules \cite{Bus} to the circle defined by $z$ yields
 \begin{equation}
 \widehat g = dx^2 + dy^2 + (d\hat z + k\,x\,dy)^2 \,,\qquad
 \widehat H = 0\,,
 \end{equation}
 which can be interpreted as a metric on the nilmanifold
 defined by $(x,y,\hat z)\sim (x+1,y,\hat z- k y)\sim (x,y+1,\hat z)\sim
(x,y,\hat z+1)$.  We would now like to apply the Buscher  rules to the 
circle defined by $y$, but it is clear from the identifications that there
is no corresponding circle action -- which is related to the fact that the nilmanifold
is not a principal $\TT^2$-bundle.
A naive application of the Buscher  rules gives results which are suspicious,
as the transformed metric does not appear to be a metric on any torus
bundle \cite{KSTT}.
The conclusion is that if one tries to T-dualize a principal torus bundle with
a non T-dualizable H-flux, in the sense of Def. \ref{defB}, it will certainly take us out
of the realm of principal torus bundles or, perhaps, there is even a genuine obstruction
for such a T-duality.   Similarly, T-duality for non principal torus bundles is not straightforward,
or perhaps problematic, for the same reasons.

\subsection{The group manifold} \label{secCB}

For an example of a T-dualizable flux consider a simple, compact Lie group 
$\sfG$.  Let $\sfT$ denote its maximal torus.  We can consider $\sfG$ as
a principal $\sfT$-bundle over the flag manifold $\sfG/\sfT$.

Let $[H_0]$ denote the generator of $H^3(\sfG,\ZZ)\cong \ZZ$, and
let $H_0$ be a de-Rham representative of $[H_0]$.  Explicitly,
if $\text{Tr}$ denotes a properly normalized trace on the Lie 
algebra $\fg$ of $\sfG$, we may take
\begin{equation}
H_0 =  \text{Tr}(\Theta\wedge\Theta\wedge\Theta)\,,
\end{equation}
where $\Theta = g^{-1}dg$ denotes the left-invariant Maurer-Cartan
1-form on $\sfG$.

Now consider the H-flux $H = k H_0$, for some positive integer $k$,
and let $X\in\ft$.
We have 
\begin{equation} \label{eqCBb}
\imath_X H = k\ \text{Tr}( X \Theta \wedge \Theta) = - k \ 
d\left( \text{Tr} ( X \Theta) \right)
\end{equation}
where, in the second step, we have used the Maurer-Cartan equation $d\Theta + 
\Theta\wedge\Theta =0$.  
First, consider $\widehat F \in \Omega^2(\sfG,\hat \ft)$ defined by
$\widehat F(X) = k\ \text{Tr}( X \Theta \wedge \Theta)$.
Since, for any $Y\in\ft$ we have $\cL_Y H=0$, the same holds for $\widehat F$, i.e.
$\cL_Y \widehat F =0$.  Moreover, for all $X,Y\in \ft$, we have
$\imath_Y \widehat F(X) = k\ \text{Tr}([X,Y]\,\Theta) = 0$ hence $\imath_Y \widehat F=0$.
This shows that $\widehat F$ is a basic, closed, $\hat\ft$-valued 2-form, i.e.
$[\widehat F]  \in H^2(\sfG/\sfT,\hat \ft)$.
Moreover, the second equality shows that we can write $\widehat F = d\widehat A$,
with $\widehat A\in\Omega^1(\widehat\sfG,\hat \ft)$ defined by
$\widehat A(X) = -k \ \text{Tr} ( X \Theta)$. 
In particular this shows that all H-fluxes on a group manifold 
are T-dualizable with respect to the maximal torus $\sfT$ (or,
with respect to any subtorus, for that matter).

Note that $\widehat A$ is {\it not} a basic 
form, but rather a globally defined connection on a T-dual $\widehat \sfT$-bundle 
$\widehat \sfG$.  It is not hard to see that, in fact, $\widehat \sfG = \sfG / (\ZZ_k)^r$,
where $r$ is the rank of $\sfG$ and $(\ZZ_k)^r \subset \sfT 
=\TT^r = \TT \times \TT \times \ldots \times \TT$  is the subgroup of $\sfT$ 
such that each $\ZZ_k\subset \TT$ with generator $\exp(2\pi i/k)$. 
In the case of $\sfG =SU(2)$, the T-dual manifold $\sfG/\ZZ_k \cong S^3/\ZZ_k$ 
is the Lens space $L(1,k)$, and we reproduce the result of \cite{MMS,BEMa}.

The example of the group manifold can be generalized to more general
principal $\sfT$-bundles over the flag manifold $\sfG/\sfT$, and leads to a whole
web of dualities with corresponding isomorphisms between the respective
twisted cohomologies and K-theories.  



\end{document}